\documentclass[12pt,preprint]{aastex}
\newcommand{\be}{\begin{equation}}
\newcommand{\ee}{\end{equation}}
\newcommand{\bbb}{\begin{eqnarray}}
\newcommand{\eee}{\end{eqnarray}}

\def\note #1]{{\bf #1]}}
\def\gwig{{\leavevmode\kern0.3em\raise.3ex\hbox{$>$}
\kern-0.8em\lower.7ex \hbox{$\sim$}\kern0.3em}}
\def\lwig{{\leavevmode\kern0.3em\raise.3ex\hbox{$<$}
\kern-0.8em\lower.7ex \hbox{$\sim$}\kern0.3em}}
\begin{document}
\title{Analytical models for Cross-correlation signal in Time-Distance Helioseismology}
\author{R. Nigam, A. G. Kosovichev and P. H. Scherrer}
\affil{W. W. Hansen  Experimental Physics Laboratory, Stanford
University, Stanford, USA}

\begin{abstract}

In time-distance helioseismology, the time signals (Doppler shifts)  at two points on the solar surface, separated by a fixed angular distance
are cross-correlated, and this leads to a wave packet signal.
Accurately measuring the travel times of these wave packets is crucial for inferring the
sub-surface properties in the Sun. The observed signal is quite noisy, and to improve the signal-to-noise ratio and make the
cross-correlation more robust, the temporal oscillation signal is phase-speed filtered at the two points in order to select waves that travel a
fixed horizontal distance. Hence a new  formula to estimate the travel times
is derived, in the presence of  a phase speed filter, and it includes both the radial and horizontal component of the oscillation displacement signal.
 It generalizes the previously used Gabor wavelet that was derived without  a phase speed  filter and included only  the radial component
of the displacement. This is important since it will  be consistent
with the observed cross-correlation that is computed using a phase speed filter,  and also
it accounts for both the components of the displacement. The new formula depends on the location of the two points on the solar
surface that are being cross correlated and  accounts for the travel time shifts at different locations on the solar surface.

\end{abstract}
\keywords{Sun: oscillations; Sun: phase speed; Sun: spherical harmonics}

\section{Introduction}
\label{sec:intro}

Time-distance helioseismology \citep{duvalletal93} constructs wave
packets by cross-correlating time signals between any two points
separated by a fixed horizontal angular distance on the solar
surface. It then measures their travel time between the two points,
by fitting a Gabor wavelet to the observed temporal
cross-correlation \citep{sashaduvall97}. This travel time is then
inverted to study different properties
\citep{kosovichev1996,zhao-thes} (e.g. sound  speed perturbations,
subsurface flows, meridional circulation) in the solar interior that
influence the wave packet, which cannot otherwise be studied by the
global oscillations. The success in applying the technique depends
on being able to design a phase speed filter of a certain width,
that selects acoustic waves within a certain  range of horizontal
phase speeds, and this improves the  signal-to-noise ratio
\citep{duvalletal97}. Waves with the same horizontal phase speed
travel the same horizontal distance on the solar surface, and
therefore can be collectively used to probe the sub-surface features
in the Sun, since they sample the same vertical depth inside the
Sun. The filtering operation is carried out in the frequency domain,
and it improves the signal-to-noise ratio by removing unwanted
signals and waves that  deviate a lot from the chosen phase speed,
as these do not contribute to the cross-correlation, but instead can
degrade it, and make the estimation of travel times inaccurate. The
phase speed filter has a Gaussian shape centered at the desired
phase speed, which is chosen to study waves traveling a particular
horizontal distance.

Previous approaches to measuring the travel time between any two points was by fitting a Gabor wavelet \citep{sashaduvall97}
to the measured cross correlation. This wavelet was derived by
 assuming that the amplitude of the solar oscillations has a 
Gaussian envelope in frequency  of a certain width and  peaked
at a  frequency where the power of the solar p-modes is
concentrated, and moreover it considered only the radial component
of the displacement. However,  during data processing a phase speed
filter is used. Also,
the observed displacement on the solar surface has both radial and
horizontal components. The horizontal component usually is ignored
in the travel-time measurements. We show that it has a
significant effect, particularly, for measurements far from the disk
center and for moderate horizontal distances, contributing to
systematic errors.

 To remove the shortcomings, we derive
a new analytical cross-correlation wavelet that incorporates the
phase speed  filter, and also includes both
the radial and horizontal components of the Doppler velocity
oscillation signal. This wavelet retains the structure of the Gabor
wavelet, and is a function of the filter parameters: central phase
speed and width,
 and also on the amplitude, phase and group travel times that depend on the oscillation
properties and the  dispersive nature of the solar medium. By including the horizontal component in the
cross-correlation we also see a dependency on the location of the two points being cross-correlated,
and moreover, the cross-correlation due to the horizontal component is a weighted sum of phase speed filtered Gabor wavelets,
with the weights depending on the location of the cross-correlated points, the horizontal travel distance
and the angular degree.

Comparing this with the original Gabor wavelet formula
\citep{sashaduvall97} we can estimate how the phase-speed filter and
the horizontal component shift the measured travel times. For the
line depth and intensity observations, that only have a scalar (radial)
component, in the weakly dispersive limit which is true for wave
packets that probe the deeper layers in the Sun and travel large
horizontal distances on the solar surface, corresponding to small
values of angular degree both formulae are similar. Hence, using the
old Gabor wavelet in this regime should not effect the travel time
measurements due to the phase speed filtering.
 On the other hand, wave packets constructed by cross-correlating time signals separated by a small horizontal distance,  are distorted by
the dispersive medium in the outer layers of the Sun, and, hence, the new formula should be used to account  for the travel time shifts due
to the phase-speed filtering procedure.

\section{Cross-correlation of  Phase Speed Filtered Signals}

\subsection{Scalar intensity and line depth filtered signal} \label{sec:analysis}

In this section, we generalize the fitting formula of
\citep{sashaduvall97} by including the phase-speed filtering. We
first consider acoustic waves, that are observed by measuring either
intensity fluctuations or line depth observations  on the solar
surface. They are represented as a sum of standing waves or normal
modes at a point $\vec r = (r, \theta, \phi)$ in the solar interior
and  time $t$, and can be written as
\begin{equation}
h(r, \theta, \phi, t) = \sum_{n,l,m} a_{nlm}\psi_{nlm}(r, \theta, \phi) \exp \left \{i(\omega_{nlm}t - \alpha_{nlm})\right \}
\label{eq1}
\end{equation}
where,  $i = \sqrt{-1}$, and  each normal mode is specified by a 3-tuple $(l, m, n)$ of integer parameters,
corresponding angular frequency $\omega_{nlm}$, the mode amplitude $a_{nlm}$, the phase $\alpha_{nlm}$
and the spatial eigenfunction $\psi_{nlm}(r, \theta, \phi)$
as a function of the radial variable $r$ and the angular variables $(\theta, \phi)$.
The integer $l$ denotes the degree and $m$ the azimuthal order,
$-l \le m \le l$, of the spherical harmonic
\begin{equation}
Y^m_l(\theta,\phi) = c_{lm} P^m_l(\cos \theta) \exp(im\phi),
\label{eq2}
\end{equation}
which is a function of the co-latitude $\theta$ and longitude $\phi$.
Here, $P^m_l$ is a Legendre function, and $c_{lm}$ is a
normalization constant. These describe the angular structure of the eigenfunctions.
The third integer $n$ of the 3-tuple $(l, m, n)$ is called the radial order.

For a spherically symmetric Sun the eigenfunctions $\psi_{nlm}(r,
\theta, \phi)$  can be separated into a radial function
$\chi_{nl}(r)$ and an angular component $Y^m_l(\theta,\phi)$
\citep{unno-book}. This representation is valid for example for the
scalar intensity observations,
\begin{equation}
\psi_{nlm}(r, \theta, \phi) = \chi_{nl}(r)Y^m_l(\theta,\phi)
\label{eq3}
\end{equation}
and also all modes with the same $n$ and $l$ have the same eigenfrequency $\omega_{nl}$,
regardless of the value of $m$. In reality the Sun is not
spherically symmetric, that causes this degeneracy in $m$ to be broken.

In time-distance helioseismology we measure the travel time of wave packets by forming a temporal
cross-correlation between the oscillation signals at two locations separated by an angular distance on the solar surface $r = R$,
where $R$ is the solar radius. To model this we represent the solar oscillations on the solar surface
as a linear superposition of normal modes, that are band-limited in angular frequency $\omega$.  
\begin{equation}
H(R, \theta, \phi, \omega) = 2\pi \sum_{n,l,m}
a_{nlm}\chi_{nl}(R)Y^m_l(\theta,\phi)\exp(-i\alpha_{nlm})\delta(\omega
- \omega_{nlm}) \label{eq4}
\end{equation}
where $H(R, \theta, \phi, \omega)$ is the temporal Fourier transform  \citep{bracewell-book}
of equation (1). We ignore mode damping and assume spherical symmetry,  hence we can  replace $\omega_{nlm}$
by $\omega_{nl}$ and invoke the band-limited nature of the solar spectrum, \citep{sashaduvall97,giles99},
 by setting  $a_{nlm}\chi_{nl}(R) = G_l(\omega)$, and we obtain from equation (\ref{eq4})
the frequency  band-limited signal $H_g(R, \theta, \phi, \omega) = H(R, \theta, \phi, \omega)G_l(\omega)$
\begin{equation}
H_g(R, \theta, \phi, \omega) = 2\pi \sum_{n,l,m}  G_l(\omega) Y^m_l(\theta,\phi)\exp(-i\alpha_{nlm})\delta(\omega - \omega_{nl})
\label{eq5}
\end{equation}
where, the Gaussian frequency function $G_l(\omega)$ captures the  band-limited nature of the amplitude of the solar modes, and is  given by
\begin{equation}
G_l(\omega) = b_l \exp\left(-\frac{(\omega - \omega_o)^2}{\delta \omega^2}\right)
\label{eq6}
\end{equation}
The  band-limited nature of the  modes is controlled by
the width $\delta \omega$,  and a central frequency $\omega_o$, where the power of the modes is peaked in the $\omega - l$ diagram.
These two parameters change for different data sets, that probe different regions of the solar interior.

A phase speed filter is applied to the signal to select  modes  from the
$\omega - l$ diagram for the purpose of constructing the cross-correlation  wave packet. 
The phase speed filter is specified by a Gaussian centered around a  phase speed $V_{ph}$ and  a
width $\delta V_{ph}$ as parameters, and is given by
\begin{equation}
F_p(V_p) = \exp\left(-\frac{\left(V_p - V_{ph}\right)^2}{\delta V_{ph}^2}\right)
\label{eq7}
\end{equation}
Where, the phase speed $V_p = \frac{\omega}{L}$,  $L = \sqrt{l(l+1)} = k_hR$, $k_h$ is the horizontal  wave number.
The role of the phase speed filter is to select waves with a small range of phase speeds, the range is
specified by the width $\delta V_{ph}$. All these waves travel almost the same horizontal distance on the solar
surface, and sample approximately  the same vertical depth
in the solar interior. Hence, it is crucial to select $\delta V_{ph}$ appropriately so as to
increase the  signal-to-noise ratio in
the cross-correlation function, and hence  be able to better resolve the sub-surface structures in the Sun.

Due to the band-limited nature of the oscillation amplitudes, only values of  $L$ which are close to $L_o = \frac{\omega_o}{V_p}$
contribute to the sum in equation (\ref{eq5}), and hence we Taylor expand $L$ about the central frequency $\omega_o$, upto first order:
\begin{equation}
L = L(\omega) = L \left [\omega_o + (\omega - \omega_o) \right] \approx L(\omega_o) +
\frac{dL}{d\omega}(\omega - \omega_o)
\label{eq8}
\end{equation}
The  equation (\ref{eq8}) can be written in terms of the angular group velocity $U_g = \frac{d\omega}{dL}$
and angular phase velocity $V_p = \frac{\omega}{L}$, evaluated at $\omega = \omega_o$, and using the fact $L(\omega_o) = \frac{\omega_o}{V_p}$, we have
\begin{equation}
L(\omega) \approx \frac{\omega}{U_g} + \left(\frac{1}{V_p} - \frac{1}{U_g}\right)\omega_o
\label{eq9}
\end{equation}

Likewise, the phase velocity $V_p(L, \omega)$ can be expanded about the point $(L_o,\omega_o)$ in the $\omega - l$ diagram to yield
\begin{equation}
F_p(L,\omega) \approx \exp\left(-\frac{V_{ph}^2\left(L - \frac{\omega}{V_{ph}}\right)^2}{\delta_f^2}\right)
\label{eq10}
\end{equation}
where, $\delta_f = \frac{\omega_o\delta V_{ph}}{V_{ph}}$, and the filter width $\delta V_{ph}$ is evaluated at $(L_o, \omega_o)$,
and is a constant.

Phase speed filtering takes place in the frequency domain and consists of multiplying the filter function $F_p(L, \omega)$
with the  band-limited  Fourier transformed  signal $H_g(R, \theta, \phi, \omega)$ from equation (\ref{eq5}).
\begin{equation}
H_{f_p} (R, \theta, \phi, \omega) = 2\pi \sum_{n,l,m} F_p(L, \omega) G_l(\omega) Y^m_l(\theta,\phi)\exp(-i\alpha_{nlm})\delta(\omega - \omega_{nl})
\label{eq11}
\end{equation}
Equation (\ref{eq11}) can be inverse Fourier transformed to yield the phase speed filtered temporal
signal $h_{f_p} (R, \theta, \phi, t)$ at a point $\vec R = (R, \theta, \phi) $ on the solar surface.
This signal now consists of a superposition of modes that lie in a region of the $\omega - l$ diagram, that is the
intersection region of the frequency band-limited Gaussian envelope of the solar modes and the phase speed  filter.
\begin{equation}
h_{f_p} (\vec R, t) = \sum_{n,l,m} F_p(L, \omega_{nl}) G_l(\omega_{nl}) Y^m_l(\theta,\phi)\exp(-i\alpha_{nlm}) \exp(i\omega_{nl}t)
\label{eq12}
\end{equation}

The temporal cross-correlation function $\psi_{f_p}(\Delta,\tau)$ of the  phase speed filtered signal $h_{f_p} (R, \theta, \phi, t)$ between
two points A and B having coordinates $\vec R_1 = (R, \theta_1, \phi_1)$ and $\vec R_2  = (R, \theta_2, \phi_2)$ respectively
on the solar surface with a fixed angular  separation
distance $\Delta$, as a function of the time lag $\tau$ is
\begin{equation}
\psi_{f_p}(\Delta,\tau) = \frac{1}{T} \int\limits_{0}^{T} h_{f_p}(\vec R_1, t)  {\bar {h}_{f_p}}(\vec R_2, t + \tau) dt
\label{eq13}
\end{equation}
where, ${\bar {h}_{f_p}}$ is the complex conjugate of the phase speed filtered signal $h_{f_p}$,
and $T$ is the length of the time series.

Substituting the expression for $h_{f_p}(\vec R, t)$ from equation (\ref{eq12}) into equation (\ref{eq13}) one gets
after applying the orthonormality of $\exp(-i\omega_{nl}t)$ in the temporal integral in equation (\ref{eq13})
\begin{equation}
\psi_{f_p}(\Delta,\tau) =  \sum_{n,l} F_p^2(L, \omega_{nl}) G_l^2(\omega_{nl}) \exp(-i\omega_{nl}\tau)
\sum_{m} \sum_{m'} Y^m_l(\theta_1,\phi_1) {\bar {Y}^{m'}}_l(\theta_2,\phi_2)
\exp \left \{-i(\alpha_{nlm} - \alpha_{nlm'})\right \}
\label{eq14}
\end{equation}
where, ${\bar {Y}^{m'}}_l(\theta_2,\phi_2)$ is the complex conjugate of $Y^{m'}_l(\theta_2,\phi_2)$.
Since the phases are random, we assume that the term $\exp \left \{-i(\alpha_{nlm} - \alpha_{nlm'})\right \}$ in the double sum for $m$ and $m'$
is zero except when $m = m'$. In this case  equation (14) becomes
\begin{equation}
\psi_{f_p}(\Delta,\tau) =  \sum_{n,l} F_p^2(L, \omega_{nl}) G_l^2(\omega_{nl}) \exp(i\omega_{nl}\tau)
\sum_{m} Y^m_l(\theta_1,\phi_1) {\bar {Y}^{m}}_l(\theta_2,\phi_2)
\label{eq15}
\end{equation}
We can simplify equation (15)  by applying the addition theorem of spherical harmonics \citep{jackson-book}
\begin{equation}
\sum_{m} Y^m_l(\theta_1,\phi_1) {\bar {Y}^{m}}_l(\theta_2,\phi_2) = \frac{(2l + 1)}{4\pi} P_l(\cos \Delta) \approx \frac{L}{2 \pi}  P_l(\cos \Delta)
\label{eq16}
\end{equation}
where, $P_l$ is the Legendre polynomial of order $l$, $L = \sqrt{l(l+1)} \approx l + \frac{1}{2}$ for large $l$ \citep{dalsgaardbook},
and $\Delta$ is the angular distance on the solar surface between the two points A $(\theta_1,\phi_1)$ and B $(\theta_2,\phi_2)$, and is
(see Appendix A equations (\ref{eq43}) and (\ref{eq44}),  Appendix D,  Figure~ \ref{fg6} and Appendix E, Figure~ \ref{fg8})
\begin{equation}
\cos \Delta = \cos \theta_1 \cos \theta_2 + \sin \theta_1 \sin \theta_2 \cos(\phi_1 - \phi_2)
\label{eq17}
\end{equation}
We can approximate for large $l$, small $\Delta$, such that $L\Delta$ is large \citep{jackson-book}
\begin{equation}
P_l(\cos \Delta) \approx J_0 \left [(2l + 1) \sin \left(\frac{\Delta}{2}\right) \right]
\approx \sqrt{\frac{2}{\pi L\Delta}} \cos \left(L\Delta - \frac{\pi}{4}\right)
\label{eq18}
\end{equation}
where, $J_0$ is the Bessel function of the first kind.
With this approximation from  equation (\ref{eq18}), the addition theorem in equation (\ref{eq16}) becomes
\begin{equation}
\sum_{m} Y^m_l(\theta_1,\phi_1) {\bar {Y}^{m}}_l(\theta_2,\phi_2)
=  \frac{2}{\sqrt{\pi\Delta}} \frac{1}{2\pi} \sqrt{\frac{L}{2}} \cos \left(L\Delta  - \frac{\pi}{4}\right)
\label{eq18a}
\end{equation}
Using  equation (\ref{eq18a}),
the equation (\ref{eq15}) for the phase speed filtered cross-correlation becomes after taking the real part
\begin{equation}
\psi_{f_p}(\Delta,\tau) =  \sum_{n,l} \frac{2}{\sqrt{\pi\Delta}} \frac{1}{2\pi} \sqrt{\frac{L}{2}}
 F_p^2(L, \omega_{nl}) G_l^2(\omega_{nl}) \cos(\omega_{nl} \tau) \cos \left(L\Delta  - \frac{\pi}{4}\right)
\label{eq19}
\end{equation}

The double sum in equation (\ref{eq19}) can be reduced to a convenient sum of integrals if we regroup the modes so that the
outer sum is over the phase speed $V_p = \frac{\omega_{nl}}{L}$, and the inner sum is over $\omega_{nl}$ \citep{sashaduvall97,giles99}.
The expression for $\psi_{f_p}(\Delta,\tau)$ is an even function of $\tau$, so we  get an identical expression
if we replace $\tau$ by $-\tau$ (negative time lag).
The product of cosines in equation (\ref{eq19}) can be transformed by using the trigonometric identity, and this results in
\begin{equation}
\psi_{f_p}(\Delta,\tau) = \sum_{V_p} \frac{1}{\sqrt{\pi\Delta}} \frac{1}{2\pi} \sqrt{\frac{L}{2}}
\sum_{\omega_{nl}} F_p^2(L, \omega_{nl}) G_l^2(\omega_{nl})
\left [ \cos \left(\omega_{nl} \tau - L\Delta + \frac{\pi}{4} \right) + \cos \left(\omega_{nl} \tau + L\Delta  - \frac{\pi}{4} \right) \right ]
\label{eq20}
\end{equation}
In equation (\ref{eq20})  we let $f_{+}(\omega_{nl} \tau) = \cos \left(\omega_{nl} \tau - L\Delta + \frac{\pi}{4} \right)$
corresponding to the positive time lag $\tau$ and $f_{-}(\omega_{nl} \tau) = \cos \left(\omega_{nl} \tau + L\Delta  - \frac{\pi}{4} \right)$
corresponding to the negative  time lag $-\tau$. Since cosine is an even function, it is seen that $f_{-}(\omega_{nl} \tau) = f_{+}(-\omega_{nl} \tau)$,
hence we can drop $f_{-}(\omega_{nl} \tau)$ from equation (\ref{eq20}), while extending the sum to negative values of $\omega_{nl}$.
We now substitute the expressions for the Gaussian envelope frequency function and the phase speed filter from equations (\ref{eq6})
and (\ref{eq10}) respectively into equation (\ref{eq20}) and since $\omega_{nl}$
is a dummy variable in the inner sum, it is replaced by $\omega$ for notational convenience. Moreover, the inner sum over $\omega_{nl}$ can be
replaced by an integral over $\omega$, hence the discrete inner sum in equation (\ref{eq20}) becomes
\begin{equation}
\psi_{f_p}(\Delta,\tau, V_p) =  \int\limits_{-\infty}^{\infty} d\omega
\exp\left(-\frac{2V_{ph}^2\left(L - \frac{\omega}{V_{ph}}\right)^2}{\delta_f^2}\right)
\exp\left [-\frac{2}{\delta \omega^2} (\omega - \omega_o)^2 \right ] \cos \left(\omega\tau - L\Delta + \frac{\pi}{4} \right)
\label{eq21}
\end{equation}
The  integral in equation (\ref{eq21}) is negligible for large frequencies hence there is very little error made in extending the
frequency limit to $\infty$ and $-\infty$.
Here, we choose the coefficient $b_l$ for the Gaussian envelope function, such that
\begin{equation}
b_l^2 = 2\pi \sqrt{\frac{2}{L}}
\label{eq21a}
\end{equation}

In order to evaluate the integral in equation (\ref{eq21}) we Taylor expand $L$, and using the 
linear dispersion relation from equation (\ref{eq9}) we get
\begin{equation}
\left(L - \frac{\omega}{V_{ph}}\right) \approx \omega\left(\frac{1}{U_g} - \frac{1}{V_{ph}}\right) -
\omega_o\left(\frac{1}{U_g} - \frac{1}{V_p}\right)
\label{eq22}
\end{equation}
\begin{equation}
(\omega\tau - L\Delta) \approx \omega\left(\tau - \tau_g \right) - \omega_o (\tau_p - \tau_g)
\label{eq23}
\end{equation}
where, the  group travel time is defined as $\tau_g = \frac{\Delta}{U_g}$ and phase travel time $\tau_p =
\frac{\Delta}{V_p}$.

Substituting  equations (\ref{eq22}) and (\ref{eq23})  into the exponent of the phase speed filter and the argument
of the cosine term respectively in equation (\ref{eq21}), and
carrying out the integral \citep{grad-ryzhik94}, which is bounded since the exponentials decay at both the limits of integration, with some algebra yields
\begin{equation}
\psi_{f_p}(\Delta,\tau,V_p) = A_{f_p}(\delta \omega,\delta_f,\tau,\tau_g,\tau_p) \cos\left \{\left( \frac{R_g(\tau_g - \tau_p)}{\tau_{ph}}
+ \varepsilon^2 \right)\frac{\omega_o(\tau - \tau_g)}{R_g^2 + \varepsilon^2} + \omega_o(\tau_g - \tau_p)  + \frac{\pi}{4} \right \}
\label{eq24}
\end{equation}
Where, $R_g = \frac{\tau_g - \tau_{ph}}{\tau_{ph}}$ and  $R_p = \frac{\tau_p - \tau_{ph}}{\tau_{ph}}$ are
dimensionless quantities representing the relative deviation of the group and phase travel times respectively from the filter phase travel time,
$\tau_{ph} =  \frac{\Delta}{V_{ph}}$ and   $\varepsilon^2 = \frac{\delta_f^2}{\delta \omega^2}$.
\begin{equation}
A_{f_p} =  \sqrt {\frac{\pi}{2}} \frac{\delta \omega ~\varepsilon }{\sqrt {R_g^2 + \varepsilon^2}}
\exp\left \{-\frac{\delta \omega^2 \varepsilon^2}{8(R_g^2 + \varepsilon^2)}\left((\tau - \tau_g)^2
 + \frac{16\omega_o^2 R_p^2 }{\delta \omega^4 \varepsilon^2}\right)\right \}
\label{eq25}
\end{equation}

The argument of the cosine term can be written as
\begin{equation}
\left( \frac{R_g(\tau_g - \tau_p)}{\tau_{ph}} + \varepsilon^2 \right)\frac{\omega_o(\tau - \tau_g)}
{R_g^2 + \varepsilon^2} + \omega_o(\tau_g - \tau_p) = \omega_{f_p}(\tau - \tau_{f_p})
\label{eq24a}
\end{equation}
where, the shifted frequency, $\omega_{f_p} = \omega_o(1 - R_{gp\varepsilon})$ and the  shifted phase travel time due to the phase speed filter is
\begin{equation}
\tau_{f_p} = \tau_p - \frac{R_{gp\varepsilon}}{1 - R_{gp\varepsilon}}(R_g - R_p)\tau_{ph}
\label{eq24b}
\end{equation}
The dimensionless parameter $R_{gp\varepsilon} = \frac{R_gR_p}{R_g^2 + \varepsilon^2}$ is zero in the absence of phase speed filtering,
which occurs when the filter width tends to infinity or in the non-dispersive limit when $R_g = R_p = 0$.
Also from equation (\ref{eq24b}) we see a shift in the phase travel times due to the phase speed filtering, which tends to zero when
$R_{gp\varepsilon}$ tends to zero. Also if we choose $V_{ph}$ to be either $V_p$ or $U_g$, then the effect of phase speed
filtering can be removed since $R_g = R_p = 0$.

Summing equation (\ref{eq24})  over phase velocities we get the final cross-correlation.
\begin{equation}
\psi_{f_p}(\Delta,\tau) = \sum_{V_p} \frac{1}{\sqrt{\pi\Delta}} \psi_{f_p}(\Delta,\tau, V_p)
\label{eq26}
\end{equation}

In the non-dispersive limit when $R_g = R_p = 0$, that is $\tau_g = \tau_p = \tau_{ph}$,
or in the limit of no phase speed filtering, large $\varepsilon$, equations (\ref{eq24}) and (\ref{eq25}) reduce to
the original Gabor wavelet formula
\begin{equation}
\psi_{f_g}(\Delta,\tau,V_p) = \delta \omega  \sqrt{\frac{\pi}{2}} \exp\left [-\frac{\delta \omega^2}{8}(\tau - \tau_g)^2\right]
\cos \left [\omega_o(\tau - \tau_p) + \frac{\pi}{4} \right]
\label{eq27}
\end{equation}
In previous papers \citep{sashaduvall97,giles99}, the factor of $\frac{\pi}{4}$ has been absorbed in the phase travel time.
By defining the phase travel time $\tau_{p}^{'} = \tau_p - \frac{\pi}{4 \omega_o}$, the Gabor wavelet in
equation (\ref{eq27}) reduces to the familiar form with the sinusoidal part being $\cos \left [\omega_o(\tau - \tau_{p}^{'}) \right]$.

We conclude that while the phase-speed filtering procedure does not
change the functional form of the basic time-distance
helioseismology fitting formula, it systematically shifts the travel
times, if the filter parameter, $V_{ph}$, is different from the
actual phase or group speeds for a given distance.

\subsection{Effect of Line-of-Sight Doppler Velocity Measurements}

We now  consider the Doppler shift observations which are commonly used in helioseismology.
The Doppler velocity depends on both  the radial and horizontal components of the displacement.
For simplicity, we take the axis of the spherical harmonics to be in the plane of the sky, orthogonal
to the line of sight.

The displacement vector $\vec d(\vec r, t)$ at location $\vec r = (r, \theta, \phi)$  and time $t$ is written
for the spherically symmetric Sun as \citep{jcd02}
\begin{equation}
\vec d(\vec r, t) =  \sum_{n,l,m} a_{nlm} \left \{ \xi_{nl}^{r}(r)Y^m_l(\theta,\phi) \hat e_r + \xi_{nl}^{h}(r)
\left [\frac{\partial Y^m_l}{\partial \theta} \hat e_{\theta} +
\frac{1}{\sin \theta} \frac{\partial Y^m_l}{\partial \phi}  \hat e_{\phi} \right] \right \} \exp \left \{i(\omega_{nl}t - \alpha_{nlm})\right \}
\label{eq28}
\end{equation}
where, $\xi_{nl}^{r}(r)$, $\xi_{nl}^{h}(r)$ are the radial and horizontal components of the displacement eigenfunctions respectively,
and $\hat e_r$, $\hat e_{\theta}$ and $\hat e_{\phi}$ are unit
vectors in $r$, $\theta$ and $\phi$ directions respectively (see Appendix A,  Figure~\ref{fg5}).

Thus, the observed Doppler signal  on the solar surface is the projection of the displacement vector onto the line-of-sight direction.
Without loss of generality we take the x-axis as the line-of-sight direction, hence the projection
is just $\vec d(\vec R, t)\cdot \hat e_x$, where $\hat e_x$ is the unit vector along the x-axis and is given by
(see Appendix A,  equation (\ref{eq40}))
\begin{equation}
\hat e_x = \hat e_r \sin \theta \cos \phi + \hat e_{\theta} \cos \theta \cos \phi - \hat e_{\phi} \sin \phi
\label{eq29}
\end{equation}
The Doppler signal on the solar surface is obtained by taking the dot product between $\vec d(\vec R, t)$ and the line of
sight direction $\hat e_x$, and is
therefore $d^x(\vec R, t) = \vec d(\vec R, t) \cdot  \hat e_x = \sum_{n,l,m} d_{nlm}^x(\vec R,t)$.
For a mode $(l, m, n)$ it is
\begin{equation}
d_{nlm}^x(\vec R) = a_{nlm} \xi_{nl}^r(R) \left \{ {Y}^{m}_l(\theta, \phi) \sin\theta \cos\phi +
\beta_{nl}(R) \left [\frac{ \partial {Y}^{m}_l(\theta, \phi) } {\partial \theta} \cos\theta \cos\phi - \frac {\sin\phi}
{\sin\theta} \frac { \partial {Y}^{m}_l(\theta, \phi)} {\partial \phi} \right ] \right \}
\label{eq30}
\end{equation}
where,
\begin{equation}
\beta_{nl}(R) = \frac{\xi_{nl}^h(R)}{\xi_{nl}^r(R)} = \frac{g(R) b_c}{R \omega_{nl}^2} \approx \frac{g(R) b_c}{R \omega_o^2 } = \frac{b_c}{\sigma^2}
\label{eq31}
\end{equation}
The ratio of horizontal to radial component $\beta_{nl}(R)$ shows dependence in frequency,
and since the solar modes are band-limited in frequency, and peaked around
$\omega_o$, we replace the frequency $\omega_{nl}$ by a constant $\omega_o = 2\pi \nu_o$. Here, $g$ is the acceleration due to gravity at the solar surface,
$\sigma$ is the dimensionless frequency and $b_c$ is a factor that depends on the boundary condition used at the solar surface $r = R$.
For the solar 5-min oscillations $\sigma^2 = \frac{R \omega_o^2}{g} \approx 1000$,
and for the boundary condition that the Lagrangian pressure perturbation  vanishes
at the solar surface: at $r = R, ~~ \delta p = 0$, leads to $b_c = 1$, and this shows that at the solar surface the radial component dominates.
For other types of boundary conditions the factor $b_c$ may depend on $L$. If one selects waves with other frequencies,
which are peaked around $\nu_1$ instead of $\nu_o =$ 3.3 mHz for the 5 minute oscillations, then the value of $\sigma^2$ at $\nu_1$ can
be easily calculated from $\sigma_1^2 \approx 1000\left(\frac{\nu_1}{3.3}\right)^2$.

With the time dependence the Doppler signal is
\begin{equation}
d_{nlm}^x(\vec R, t) = d_{nlm}^x(\vec R) \exp \left \{i(\omega_{nl}t - \alpha_{nlm})\right \}
\label{eq32}
\end{equation}

The cross-correlation $\psi_{f_p}^{d^x}(\vec R_1, \vec R_2, \tau)$ can be computed for the
phase speed filtered Doppler signal
$d_{f_p}^x (R, \theta, \phi, t) = \sum_{n,l,m} F_p(L, \omega_{nl}) d_{nlm}^x(\vec R, t)$
in a similar manner as for equation (\ref{eq12}) and involves
the  product of the projected line of sight Doppler signals  at the two locations
A $(\theta_1,\phi_1)$ and B $(\theta_2,\phi_2)$ on the solar surface (See Figures~ \ref{fg6} and \ref{fg8}).
Here we have replaced $a_{nlm} \xi_{nl}^r(R)$ by the Gaussian frequency function $G_l(\omega)$, which models the amplitude of the solar modes.
\begin{equation}
\psi_{f_p}^{d^x}(\vec R_1, \vec R_2, \tau) = \frac{1}{T} \int\limits_{0}^{T} d_{f_p}^x(\vec R_1, t){\bar d_{f_p}^x}(\vec R_2, t + \tau) dt
\label{eq33}
\end{equation}
In evaluating this cross-correlation we need the summation over $m$ of the spherical harmonics and its derivatives,
which are given in Appendix C.

Collecting all the sums over $m$ we define the total sum (see Appendix C equations (\ref{eq89})-(\ref{eq92}))
\begin{equation}
d_{nl}^x(\vec R_1, \vec R_2) = \sum_{m} d_{nlm}^x(\vec R_1)d_{nlm}^x(\vec R_2) =
f_0 P_l(\cos\Delta) + f_1 \frac {dP_l(\cos\Delta)} {d\Delta} + f_2 \frac {d^2P_l(\cos\Delta)} {d\Delta^2}
\label{eq34}
\end{equation}

Substituting the expressions for $P_l(\cos\Delta)$ and its derivatives
from Appendix B equations (\ref{eq50}), (\ref{eq51}) and (\ref{eq52}), we obtain for large $l$, small $\Delta$, such that $L\Delta$ is large
\begin{equation}
d_{nl}^x(\vec R_1, \vec R_2)
=  \sqrt{\frac{2}{\pi L\Delta}} \left [f_0 \cos \left(L \Delta - \frac{\pi}{4} \right)
- lf_1 \sin \left(L \Delta  - \frac{\pi}{4} \right) - l^2 f_2 \cos \left(L \Delta  - \frac{\pi}{4} \right) \right ]
\label{eq35}
\end{equation}
where $L = \sqrt{l(l+1)}$ and $l = (L -\frac{1}{2})$.
Equation (\ref{eq35}) can be written as
\begin{equation}
d_{nl}^x(\vec R_1, \vec R_2) =  \sqrt{\frac{2}{\pi L\Delta}} \left [A_l \cos \left(L \Delta - \frac{\pi}{4} \right) -
B_l \sin \left(L \Delta - \frac{\pi}{4} \right) \right ] =
 C_l \sqrt{\frac{2}{\pi L\Delta}} \cos \left(L \Delta - \frac{\pi}{4} + \zeta \right)
\label{eq36}
\end{equation}
where, the phase factor $\zeta(\theta_1, \phi_1, \theta_2, \phi_2) = \tan^{-1} \left (B_l/A_l \right)$,
$A_l = (f_0 - l^2 f_2)$, $B_l = lf_1$ and $C_l = \sqrt{A_l^2 + B_l^2}$.
The value of $L$ outside the trigonometric functions is evaluated at $\omega_o$ and is
$L(\omega_o) = \frac{\omega_o}{V_p} = \frac{\omega_o \tau_p}{\Delta}$. The phase factor $\zeta$ in general is a function of
frequency $\omega_{nl}$, but since the solar oscillations are band-limited and their power is peaked at the frequency $\omega_o$,
$\zeta$ is evaluated at the central frequency $\omega_o$.
This also simplifies the evaluation of the integral, and a closed form expression can be derived.
The phase shift $\zeta$ is measured in radians but can be converted to units of time by dividing
by $\omega_o$.

The cross-correlation is
\begin{equation}
\psi_{f_p}^{d^x}(\vec R_1, \vec R_2, \tau) = \sum_{n,l} F_p^2(L, \omega_{nl}) G_l^2(\omega_{nl}) \cos(\omega_{nl}\tau) d_{nl}^x(\vec R_1, \vec R_2)
\label{eq37}
\end{equation}
Substituting the expression for $d_{nl}^x(\vec R_1, \vec R_2)$ from equation (\ref{eq36}) into equation (\ref{eq37}) and
transforming the
product of the cosines to a sum we obtain
\begin{equation}
\psi_{f_p}^{d^x}(\vec R_1, \vec R_2, \tau) = \sum_{V_p} \frac{2 \pi C_l}{L \sqrt{\pi\Delta}} \frac{1}{2\pi} \sqrt{\frac{L}{2}}
\sum_{\omega_{nl}} F_p^2(L, \omega_{nl}) G_l^2(\omega_{nl})
\left [ f_{+h}(\omega_{nl} \tau, \zeta) + f_{-h}(\omega_{nl} \tau, \zeta) \right]
\label{eq38}
\end{equation}
where, $f_{+h}(\omega_{nl} \tau, \zeta) =  \cos \left(\omega_{nl} \tau - L\Delta + \frac{\pi}{4} - \zeta \right)$ and
$f_{-h}(\omega_{nl} \tau, \zeta) = \cos \left(\omega_{nl} \tau + L\Delta  - \frac{\pi}{4} + \zeta \right)$ correspond
to positive and negative time lags respectively.
Evaluating the inner sum in equation (\ref{eq38})  in a similar manner as in equation (\ref{eq20}) one obtains

\begin{equation}
\psi_{f_p}^{d^x}(\vec R_1, \vec R_2, \tau, V_p) = A_{f_p}(\delta \omega,\delta_f,\tau,\tau_g,\tau_p)
\cos \left \{\omega_{f_p}(\tau - \tau_{f_{ph}}) + \frac{\pi}{4} \right \}
\label{eq39}
\end{equation}

The shifted phase travel  time $\tau_{f_{ph}}$ due to the horizontal component is therefore,
\begin{equation}
\tau_{f_{ph}} = \tau_{f_p} + \frac{\zeta}{\omega_{f_p}}
\label{eq39a}
\end{equation}

\begin{figure}[!ht]
\epsscale{0.8} \plotone{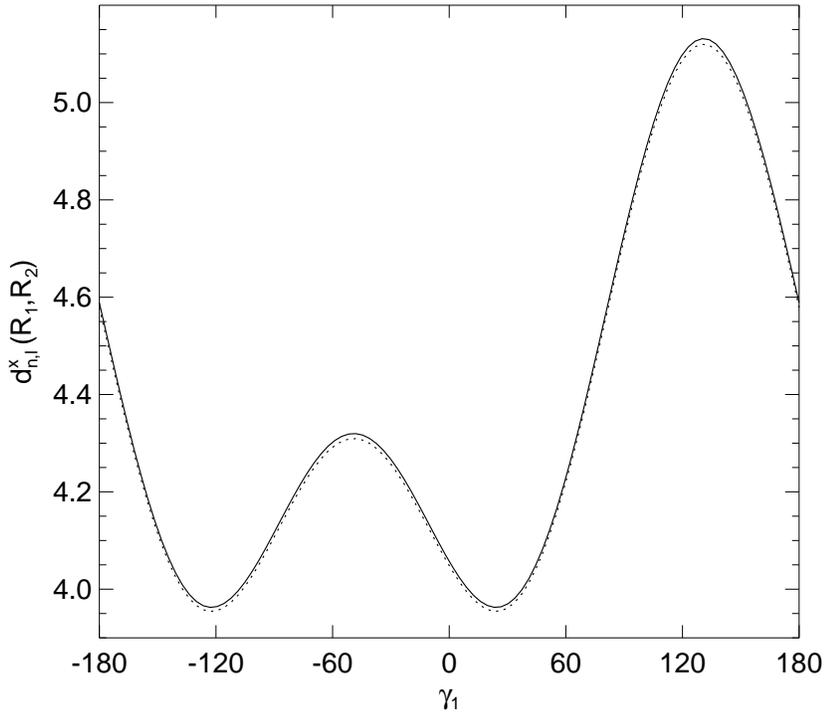} \caption{Comparison of function
$d_{nl}^x(\vec R_1,\vec R_2)$, calculated numerically by direct mode
summation given by the middle term in equation~\ref{eq34} (solid curve)
and the analytical expression equation~\ref{eq36} (dotted curve). In this
example, the central $\vec R_1$ point is located at
$\theta_1=60^\circ$ and $\phi_1=30^\circ$, point $\vec R_2$ is
located at a distance $\Delta=8.4^\circ$ from $\vec R_1$, and the
angle, $\gamma_1$
 between the directions
from $\vec R_1$ to the North pole and point $\vec R_2$ varies from $-180^\circ$ to $180^\circ$,
as illustrated in Figure~\ref{fg6}, Appendix D.} \label{fg1}
\end{figure}

Summing equation (\ref{eq39})  over phase velocities we get the final cross-correlation.
\begin{equation}
\psi_{f_p}^{d^x}(\vec R_1, \vec R_2, \tau) =
\sum_{V_p} \frac{2 \pi C_l}{L \sqrt{\pi\Delta}} \psi_{f_p}^{d^x}(\vec R_1, \vec R_2,\tau, V_p)
\label{eq39b}
\end{equation}
In deriving this equation we used the asymptotic formula for the
Legendre functions, which allowed us to obtain the explicit
expression for the phase shift. We verified by direct numerical
calculations of the cross-covariance functions that this
approximation is sufficiently accurate. A comparison of the analytical
formula in equation (\ref{eq36}) for $d_{nl}^x(\vec R_1,\vec R_2)$  with the direct numerical
mode summation is shown in Figure~\ref{fg1}.
This approximation holds
when the travel times are much greater than the oscillation period of five minutes,
which is the case for typical time-distance measurements.

From the  equations (\ref{eq35}) and (\ref{eq37}) we observe that the cross-correlation
of the Doppler line of sight velocity involving the horizontal component of the displacement
is a weighted sum of the phase speed filtered Gabor wavelets, with the weights depending on the location
$(\theta_1, \phi_1, \theta_2, \phi_2)$ of the two
points being cross-correlated, and the angular degree $l$ of the modes. This
weighted sum can be conveniently combined incorporating a phase factor into the cosine term.

From equation (\ref{eq39a}) we see a shift in phase travel times due
to the horizontal component in addition to that introduced by the
phase speed filtering.  In Figure~\ref{fg2}, we see a variation in
the travel time shift due to the horizontal component as we move
point B (see Appendix E, Figure~ \ref{fg8}) around the annulus by
changing the angle $\gamma_1$ (see  Appendix D, Figure~\ref{fg6}), cross
correlating with the fixed point A. Figure~\ref{fg3} illustrates the
phase shift for the projection on the solar disk. The greatest phase
shift changes happen for the outmost points in the direction from
the disk center.

Averaging the travel time shift over the annular angle $\gamma_1$,
for different horizontal distances $\Delta$ and distances from disk
center $\eta$, to simulate the observed mean travel times, we show
in Figure~\ref{fg4} that this time shift increases (becomes more
negative) as we go away from disk center and also  for increasing
horizontal distances. The shifts are quite appreciable, on the order of a few 
seconds,  and have been observed in the real data from SOHO/MDI
\citep{duvall02}. At disk center C, (see Appendix E, Figure~\ref{fg7}) which
according to our convention is at $(y, z) = (0, 0)$, which is the
origin of the YZ plane perpendicular to the line of sight direction
$\hat e_x$. In polar coordinates the disk center is specified by
$(\theta, \phi) = (\frac{\pi}{2}, 0)$, at which the horizontal
component is zero in equation (\ref{eq30}), and the radial component
is maximum. We see from Figure~\ref{fg2}, when point A is at disk
center, the time shift is independent of $\gamma_1$, moreover there
will always be a small travel time shift for this case since point B
is at a distance $\Delta$ from the disk center, and it has a
non-zero horizontal component that when cross-correlated with the
radial component of the signal at point A, which is at disk center,
leads to a travel time shift.

\begin{figure}[!ht]
\epsscale{0.8} \plotone{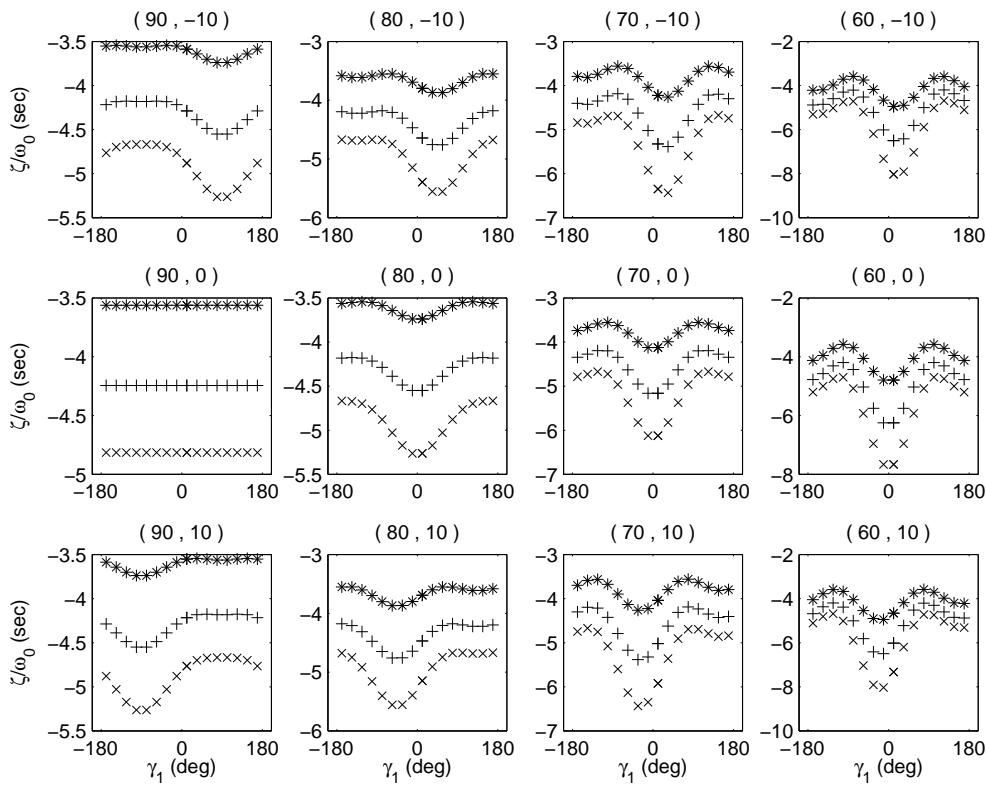} \caption{Phase shift (Y axis)
$\frac{\zeta}{\omega_o}$ in seconds as a function of  angle
$\gamma_1$ (X axis) in degrees. The top most row corresponds to
$\phi_1 = -10$ degrees, the next rows are in increments of 10
degrees. The left most column corresponds to $\theta_1 = 90$
degrees, the other columns are in decrements of $10$ degrees. With
this convention the  left most plot of the second row corresponds to
$(\theta_1, \phi_1) = (90^\circ, 0^\circ)$, for the annulus central
point A, which is at disk center C. The star (*), plus (+) and cross
(x) lines respectively in each plot correspond to $\Delta =$ 9.84,
15.36 and 20.64  degrees, and $\tau_p = $ 58.89, 69.24 and 77.01
minutes respectively. In this figure there is no shift due to phase
speed filtering.} \label{fg2}
\end{figure}

\begin{figure}[!ht]
\epsscale{0.5} \plotone{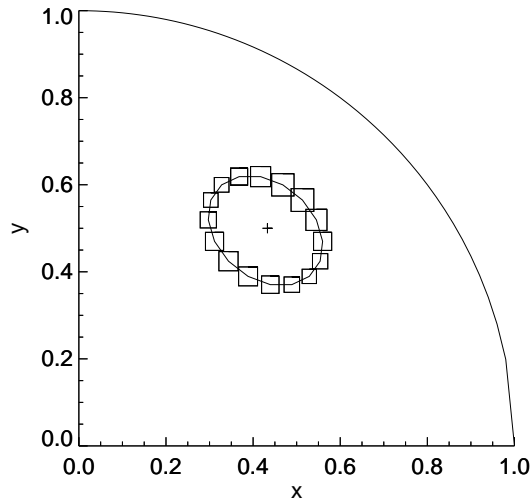} \caption{Illustration of the phase
shift variations for a projection on the solar disk. The travel
times are measured between the central point, located at
$\theta_1=60$ degrees and $\phi_1=30$ degrees, and surrounding
points at distance $\Delta=8.4$ degrees and corresponding $\tau_p =
$ 55.59 minutes. The size of the symbols is proportional to the
magnitude of the phase shifts which varies from -0.07 to -0.12 radians
(corresponding to travel time shifts from -3.5 seconds to -5.5 seconds).
In this Figure there is no  phase speed filtering.}
\label{fg3}
\end{figure}

\begin{figure}[!ht]
\epsscale{0.8} \plotone{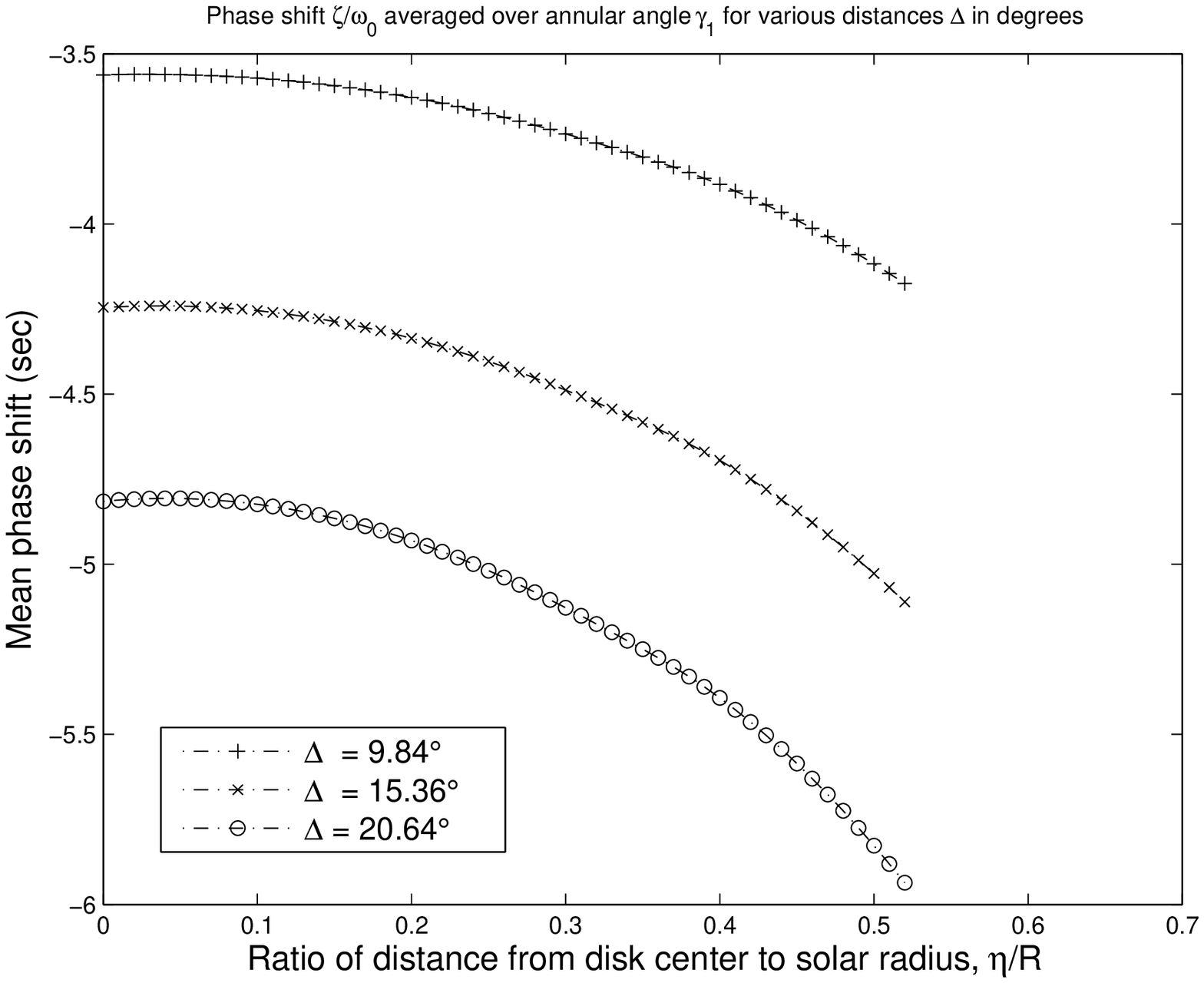} \caption{The mean phase shift in
seconds (Y axis) got by averaging $\frac{\zeta}{\omega_o}$  over the
annular angle $\gamma_1$ as a function of  ratio with respect to
solar radius of distance of the central point A from disc center C,
$\frac{\eta}{R}$ (X axis). The star (*), plus (+)  and cross (x)
lines respectively correspond to $\Delta =$ 9.84, 15.36 and 20.64
degrees, and $\tau_p = $ 58.89, 69.24 and 77.01 minutes
respectively. We observe a maximum travel time shift of about 6
seconds. This plot has no phase speed filtering, whose effect can be
removed by choosing $V_{ph}$ appropriately (see Sec.2.1).}
\label{fg4}
\end{figure}

\section{Dispersive and Non-Dispersive Wave packets: Relation to the Gabor Wavelet without Phase Speed  Filtering}

A dispersive medium is where the phase velocity $V_p$ is quite
different from the group velocity $U_g$
\citep{lighthill-book,whitham-book}. Due to this for a fixed
distance $\Delta$, the group and phase travel times are different.
In the solar case the wave packets that travel short distances,
probe the outer layers of the Sun, which are quite dispersive due to
the presence of large gradients. Hence the wave packets got by the
cross-correlation in equations (\ref{eq24}) and (\ref{eq25})   are
distorted as they travel through the medium. Moreover, the
approximations made by retaining only the first order terms in the
Taylor expansion of the dispersion relation in equation (\ref{eq8}) may be inadequate to
completely model the physical effects associated with small
distances, hence care must be exercised in interpreting the results
got by using these wavelets for small distances.

On the other hand, wave packets that travel large angular distances $\Delta$, probe the deeper layers of the Sun,
that are weakly dispersive, and hence undergo less distortion. It is quite interesting to note
that the Gabor wavelet in equation (\ref{eq27}),  in the absence of the phase speed  filter is got in the non-dispersive limit
of equations (\ref{eq24}) and (\ref{eq25}) when $R_g = R_p = 0$, that is $\tau_g = \tau_p = \tau_{ph}$, for the
intensity or line depth signal which has only the radial component.
This is quite intuitive since in this limit there is no phase speed filtering, and $F_p(L, \omega) = 1$, hence the phase speed
filtered signal in equation (\ref{eq11}) reduces to the  frequency band-limited signal in equation (\ref{eq5}). This is consisted both physically
and mathematically. To form wave packets the medium need not be dispersive. Waves in a range of frequencies and wave lengths
can interfere to form a wave packet. The wave packets that travel through a dispersive medium are distorted,
where as those through a non-dispersive medium  are not distorted, but  retain their original shape.

The dimensionless quantities $R_p$ and $R_g$ measure the amount of deviation from non-dispersiveness.
Hence to fit the observed temporal cross-correlation for the intensity or line
depth signal where, $R_g$ and $R_p$ are small, there will be little
error in using the  Gabor wavelet of equation (\ref{eq27}), since the new Gabor wavelet of equations (\ref{eq24}) and (\ref{eq25}) is  very close
to the old Gabor wavelet \citep{sashaduvall97,giles99} in this non-dispersive or weakly dispersive limit
for the line depth or scalar intensity signal that have only the radial component.

\section{Conclusion}
In this paper we introduce a mathematical formalism to deal with the horizontal component
of the displacement, which is used to derive a new  formula
to estimate travel times in time-distance helioseismology.
It models the observed cross-correlation, by considering both the
radial and horizontal components of the displacement, and also includes the  phase speed filter on
the frequency band-limited solar oscillations.
The form of the Gabor wavelet is retained by the present formula, and in the non-dispersive limit
for the line depth or intensity observations it reduces
to the previously used Gabor wavelet that was derived without a phase speed filter, considering only the radial component.
Including the horizontal component, the cross-correlation is a weighted sum of phase speed filtered Gabor wavelets,
with the weights depending on the angular position of the two points being cross-correlated,
and the angular degree of the modes.
Due to the phase speed filtering and inclusion of the horizontal
component of the displacement, shifts in phase travel times are observed. In particular, the horizontal component
induces a systematic shift in travel times, whose absolute value increases as we go away from disk center and also
for large  horizontal distances.
This will in turn have implications
in the inversions to study flows and the  sub-surface properties in the Sun and also to formulate other local
helioseismic approaches that involve the horizontal component.

\section{Acknowledgments}

This project was supported by the NASA MDI grant to Stanford
University.

\section{Appendix A}

\begin{figure}[!ht]
\epsscale{0.8} \plotone{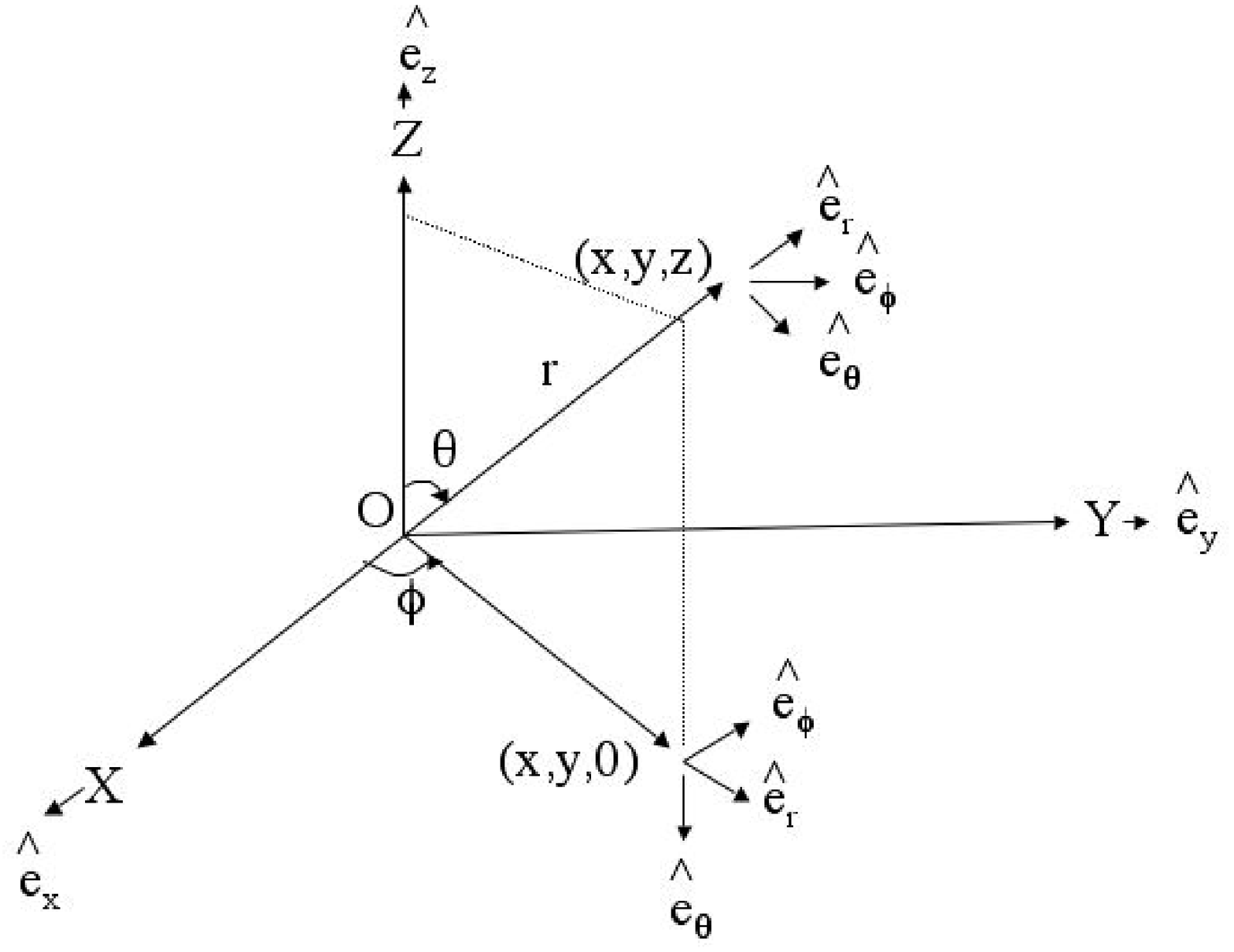} \caption{Spherical Coordinate
System. The line of sight direction is along $\hat e_x$.}
\label{fg5}
\end{figure}

We define the spherical coordinate system used in this paper
(see Figure~\ref{fg5}).

Consider fixed cartesian unit vectors $(\hat e_x, \hat e_y, \hat e_z)$ and the corresponding spherical
unit vectors $(\hat e_r, \hat e_{\theta}, \hat e_{\phi})$ that vary with the angles $(\theta, \phi)$.
The equations relating these sets of unit vectors are

\begin{equation}
\hat e_x = \hat e_r \sin \theta \cos \phi + \hat e_{\theta} \cos \theta \cos \phi - \hat e_{\phi} \sin \phi
\label{eq40}
\end{equation}
\begin{equation}
\hat e_y = \hat e_r \sin \theta \sin \phi + \hat e_{\theta} \cos \theta \sin \phi + \hat e_{\phi} \cos \phi
\label{eq41}
\end{equation}
\begin{equation}
\hat e_z = \hat e_r \cos \theta - \hat e_{\theta} \sin \theta
\label{eq42}
\end{equation}

\begin{figure}[!ht]
\epsscale{0.8} \plotone{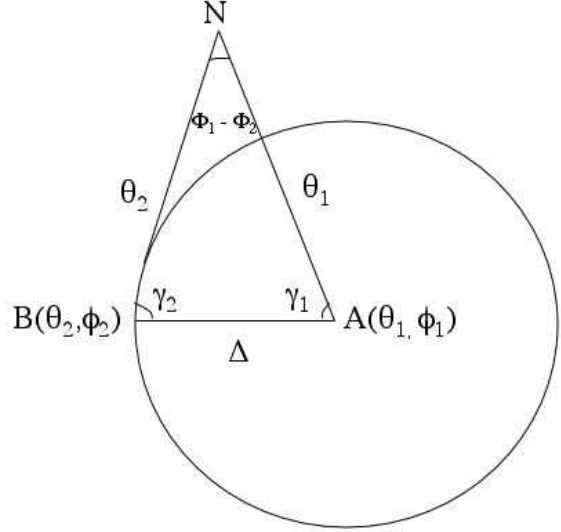} \caption{The Spherical Triangle
showing $\Delta$, which is the angular separation of the great
circle arc between points A and B on the solar surface. Here, N is
the North pole.} \label{fg6}
\end{figure}

Any point $(x, y, z)$ in spherical coordinates $(r, \theta, \phi)$ is
$x = r \sin \theta \cos \phi$, $y = r\sin \theta \sin \phi$ and $z = r \cos \theta$.
Consider two points A ~~$(\theta_1, \phi_1)$ and B ~~ $(\theta_2, \phi_2)$ on the solar surface $r = R$ (see Figure~\ref{fg6}, Appendix D),
with coordinates $\vec R_1 = (x_1, y_1, z_1)$ and $\vec R_2 = (x_2, y_2, z_2)$, the angular distance $\Delta$ between them
can be calculated by taking the dot product of the unit vectors at these two locations and is given by
\begin{equation}
\cos \Delta = \frac {(\vec R_1 \cdot \vec R_2)}{R^2} = \frac{x_1x_2 + y_1y_2 + z_1z_2}{R^2}
\label{eq43}
\end{equation}
where, $(x_1, y_1, z_1) = (R \sin \theta_1 \cos \phi_1, R \sin \theta_1 \sin \phi_1, R \cos \theta_1)$ and \\
$(x_2, y_2, z_2) = (R \sin \theta_2 \cos \phi_2 , R \sin \theta_2
\sin \phi_2, R \cos \theta_2)$ (see Figure~\ref{fg6}, Appendix D). Substituting
this into equation (\ref{eq43}) we obtain
\begin{equation}
\cos \Delta = \cos \theta_1 \cos \theta_2 + \sin \theta_1 \sin \theta_2 \cos(\phi_1 - \phi_2)
\label{eq44}
\end{equation}

\section{Appendix B}

We derive various approximations for $P_l(\cos \Delta)$ and its derivatives for
large $l$, small $\Delta$ and large $L\Delta$ following the approach of \citep{jackson-book}.
\begin{equation}
P_l(\cos \Delta) \approx J_0 \left [(2l + 1) \sin \left(\frac{\Delta}{2}\right) \right]
\approx \sqrt{\frac{2}{\pi L\Delta}} \cos \left(L\Delta - \frac{\pi}{4}\right)
\label{eq45}
\end{equation}
From the recursion relation \citep{jackson-book},
\begin{equation}
\sin\Delta \frac {dP_l(\cos\Delta)} {d\Delta} - l\cos\Delta  P_l(\cos\Delta) + lP_{l-1} (\cos\Delta) = 0
\label{eq46}
\end{equation}
For small $\Delta$  equation (\ref{eq46}) becomes,
\begin{equation}
\Delta \frac {dP_l(\cos\Delta)} {d\Delta} - lP_l(\cos\Delta) + lP_{l-1} (\cos\Delta) = 0
\label{eq47}
\end{equation}
Differentiating equation (\ref{eq46}) with respect to $\Delta$ gives
\begin{equation}
\sin\Delta \frac {d^2P_l(\cos\Delta)} {d\Delta^2} - (l-1) \cos\Delta \frac {dP_l(\cos\Delta)} {d\Delta} + l \frac {dP_{l-1} (\cos\Delta)} {d\Delta} + l\sin\Delta P_l(\cos\Delta) = 0
\label{eq48}
\end{equation}
For small $\Delta$ and large $l$, equation (\ref{eq48}) becomes,
\begin{equation}
\Delta \frac {d^2P_l(\cos\Delta)} {d\Delta^2} - l \frac {dP_l(\cos\Delta)} {d\Delta} + l \frac {dP_{l-1} (\cos\Delta)} {d\Delta} + l\Delta P_l(\cos\Delta) = 0
\label{eq49}
\end{equation}
Now for small $\Delta$, large $l$ and large $L\Delta$ the various approximations are after using the recursion relations from equations (\ref{eq47}) and
(\ref{eq49})
\begin{equation}
P_l(\cos \Delta) \approx \sqrt{\frac{2}{\pi L\Delta}}  \cos \left(L\Delta - \frac{\pi}{4}\right)
\label{eq50}
\end{equation}

\begin{equation}
 \frac {dP_l(\cos\Delta)} {d\Delta} \approx - l \sqrt{\frac{2}{\pi L\Delta}} \sin \left(L\Delta - \frac{\pi}{4}\right)
\label{eq51}
\end{equation}

\begin{equation}
\frac {d^2P_l(\cos\Delta)} {d\Delta^2} \approx - l^2 \sqrt{\frac{2}{\pi L\Delta}}  \cos \left(L\Delta - \frac{\pi}{4}\right)
\label{eq52}
\end{equation}

\section{Appendix C}

Using the addition theorem of the derivatives of  spherical harmonics \citep{winch-roberts95}, we derive the expressions
for the terms in equation (\ref{eq34}).

The angular distance $\Delta$ is,
\begin{equation}
\cos \Delta = \cos \theta_1 \cos \theta_2 + \sin \theta_1 \sin \theta_2 \cos(\phi_1 - \phi_2)
\label{eq53}
\end{equation}

Differentiating equation (\ref{eq53})
with respect to $\theta_1$, $\theta_2$, $\phi_1$ and $\phi_2$ we obtain

\begin{equation}
\frac{\partial \Delta}{\partial \theta_1}  = \frac{\sin \theta_1 \cos \theta_2 - \cos \theta_1 \sin \theta_2 \cos(\phi_1 - \phi_2)}{\sin \Delta}
= \cos \gamma_1
\label{eq54}
\end{equation}

\begin{equation}
\frac{\partial \Delta}{\partial \theta_2} = \frac{\cos\theta_1 \sin\theta_2 - \sin\theta_1 \cos\theta_2 \cos(\phi_1 - \phi_2)}{\sin \Delta}
= \cos \gamma_2
\label{eq55}
\end{equation}

\begin{equation}
\frac{\partial \Delta}{\partial \phi_1}  = \frac{\sin \theta_1 \sin\theta_2 \sin(\phi_1 - \phi_2)}{\sin \Delta} = \sin \theta_1 \sin \gamma_1
\label{eq56}
\end{equation}

\begin{equation}
\frac{\partial \Delta}{\partial \phi_2}  = - \frac{\sin\theta_1 \sin \theta_2 \sin(\phi_1 - \phi_2)}{\sin \Delta} = - \sin \theta_2 \sin \gamma_2
\label{eq57}
\end{equation}

The expressions in equations (\ref{eq54})-(\ref{eq57}) can be
equated to sine and cosine functions of $\gamma_1$ and $\gamma_2$
\citep{winch-roberts95}, which are shown in Figure~\ref{fg6}, Appendix D using
the results of spherical trigonometry \citep{smart-book}.

According to the addition theorem of spherical harmonics \citep{jackson-book}
\begin{equation}
s_{nl}^{rr} = \sum_{m} Y^m_l(\theta_1,\phi_1) {\bar {Y}^{m}}_l(\theta_2,\phi_2) = \frac{(2l + 1)}{4\pi} P_l(\cos \Delta)
\label{eq58}
\end{equation}

For notational convenience we denote $N_l =  \frac{(2l + 1)}{4\pi}$.
Differentiating equation (\ref{eq58}) with respect to $\theta_2$ and applying the chain rule leads to
\begin{equation}
s_{nl}^{r \theta} = \sum_{m} Y^m_l(\theta_1,\phi_1) \frac{ \partial {\bar {Y}^{m}}_l(\theta_2,\phi_2)}{\partial \theta_2} =
N_l \frac {dP_l(\cos\Delta)} {d \Delta} \frac{\partial \Delta}{\partial \theta_2}
\label{eq59}
\end{equation}

Differentiating equation (\ref{eq58}) with respect to $\phi_2$ and applying the chain rule leads to
\begin{equation}
s_{nl}^{r \phi} = \sum_{m} Y^m_l(\theta_1,\phi_1) \frac{ \partial {\bar {Y}^{m}}_l(\theta_2,\phi_2)}{\partial \phi_2} =
N_l \frac {dP_l(\cos\Delta)} {d \Delta} \frac{\partial \Delta}{\partial \phi_2}
\label{eq60}
\end{equation}

Differentiating equation (\ref{eq58}) with respect to $\theta_1$ and applying the chain rule leads to
\begin{equation}
s_{nl}^{\theta r} = \sum_{m} \frac{ \partial {Y}^{m}_l(\theta_1,\phi_1)}{\partial \theta_1} Y^m_l(\theta_2,\phi_2) =
N_l \frac {dP_l(\cos\Delta)} {d \Delta} \frac{\partial \Delta}{\partial \theta_1}
\label{eq61}
\end{equation}

Differentiating equation (\ref{eq61}) with respect to $\theta_2$ and applying the product and chain rule gives
\begin{equation}
s_{nl}^{\theta \theta} = \sum_{m} \frac{ \partial {Y}^{m}_l(\theta_1,\phi_1)}{\partial \theta_1} \frac{ \partial {\bar {Y}^{m}}_l(\theta_2,\phi_2)}{\partial \theta_2}
 = N_l \left [\frac{\partial \Delta} {\partial \theta_1}  \frac{\partial \Delta} {\partial \theta_2} \frac {d^2 P_l(\cos\Delta)} {d\Delta^2}
+ \frac{\partial }{\partial \theta_2} \left(\frac{\partial \Delta} {\partial \theta_1} \right)  \frac {dP_l(\cos\Delta)} {d\Delta} \right ]
\label{eq62}
\end{equation}

Differentiating equation (\ref{eq61}) with respect to $\phi_2$ and applying the product and chain rule gives
\begin{equation}
s_{nl}^{\theta \phi} = \sum_{m} \frac{ \partial {Y}^{m}_l(\theta_1,\phi_1)}{\partial \theta_1} \frac{ \partial {\bar {Y}^{m}}_l(\theta_2,\phi_2)}{\partial
\phi_2} = N_l \left [\frac{\partial \Delta} {\partial \theta_1}  \frac{\partial \Delta} {\partial \phi_2} \frac {d^2 P_l(\cos\Delta)} {d\Delta^2}
+ \frac{\partial }{\partial \phi_2} \left(\frac{\partial \Delta} {\partial \theta_1} \right)  \frac {dP_l(\cos\Delta)} {d\Delta} \right ]
\label{eq63}
\end{equation}

Differentiating equation (\ref{eq58}) with respect to $\phi_1$ and applying the chain rule leads to
\begin{equation}
s_{nl}^{\phi r} = \sum_{m} \frac{ \partial {Y}^{m}_l(\theta_1,\phi_1)}{\partial \phi_1} {\bar {Y}^{m}}_1(\theta_2,\phi_2) =
N_l \frac {dP_l(\cos\Delta)} {d\Delta} \frac{\partial \Delta}{\partial \phi_1}
\label{eq64}
\end{equation}

Differentiating equation (\ref{eq64}) with respect to $\theta_2$ and applying the product and chain rule gives
\begin{equation}
s_{nl}^{\phi \theta} = \sum_{m} \frac{ \partial {Y}^{m}_l(\theta_1,\phi_1)}{\partial \phi_1} \frac{ \partial {\bar {Y}^{m}}_l(\theta_2,\phi_2)}{\partial \theta_2}
 = N_l \left [\frac{\partial \Delta} {\partial \phi_1}  \frac{\partial \Delta} {\partial \theta_2} \frac {d^2 P_l(\cos\Delta)} {d\Delta^2}
+ \frac{\partial }{\partial \theta_2} \left(\frac{\partial \Delta} {\partial \phi_1} \right)  \frac {dP_l(\cos\Delta)} {d\Delta} \right ]
\label{eq65}
\end{equation}

Differentiating equation (\ref{eq64}) with respect to $\phi_2$ and applying the product and chain rule gives
\begin{equation}
s_{nl}^{\phi \phi} = \sum_{m} \frac{ \partial {Y}^{m}_l(\theta_1,\phi_1)}{\partial \phi_1} \frac{ \partial {\bar {Y}^{m}}_l(\theta_2,\phi_2)}{\partial \phi_2}
 = N_l \left [\frac{\partial \Delta} {\partial \phi_1}  \frac{\partial \Delta} {\partial \phi_2} \frac {d^2 P_l(\cos\Delta)} {d\Delta^2}
+ \frac{\partial }{\partial \phi_2} \left(\frac{\partial \Delta} {\partial \phi_1} \right)  \frac {dP_l(\cos\Delta)} {d\Delta} \right ]
\label{eq66}
\end{equation}

Now using equations (\ref{eq30}),  (\ref{eq58})-(\ref{eq66}) we obtain for the various terms in equation (\ref{eq34})

\begin{equation}
s^1_{nl} = \sin\theta_1 \cos\phi_1 \sin \theta_2 \cos \phi_2 s_{nl}^{rr} = f^{rr}_{10}(\theta_1,\phi_1,\theta_2,\phi_2) P_l(\cos\Delta)
\label{eq67}
\end{equation}

For notational convenience we denote the angular coordinates of points $\vec R_1$ and $\vec R_2$ by $\vec a_{12} = (\theta_1,\phi_1,\theta_2,\phi_2)$

\begin{equation}
s^2_{nl} = \beta_{nl}(R) \sin\theta_1 \cos\phi_1 \cos\theta_2 \cos\phi_2  s_{nl}^{r \theta} =
f^{r \theta}_{21}(\vec a_{12}) \frac {dP_l(\cos\Delta)} {d\Delta}
\label{eq68}
\end{equation}

\begin{equation}
s^3_{nl} = - \beta_{nl}(R) \sin\theta_1 \cos\phi_1 \frac {\sin\phi_2} {\sin\theta_2} s_{nl}^{r \phi} =
f^{r \phi}_{31}(\vec a_{12}) \frac {dP_l(\cos\Delta)} {d\Delta}
\label{eq69}
\end{equation}

\begin{equation}
s^4_{nl} = \beta_{nl}(R) \cos\theta_1 \cos\phi_1 \sin\theta_2 \cos\phi_2 s_{nl}^{\theta r}  =
f^{\theta r}_{41}(\vec a_{12}) \frac {dP_l(\cos\Delta)} {d\Delta}
\label{eq70}
\end{equation}

\begin{equation}
s^5_{nl} = [\beta_{nl}(R)]^2 \cos\theta_1 \cos\phi_1 \cos\theta_2 \cos\phi_2 s_{nl}^{\theta \theta} =
f^{\theta \theta}_{52}(\vec a_{12}) \frac {d^2P_l(\cos\Delta)} {d\Delta^2} + f^{\theta \theta}_{51}(\vec a_{12}) \frac {dP_l(\cos\Delta)} {d\Delta}
\label{eq71}
\end{equation}

\begin{equation}
s^6_{nl} = - [\beta_{nl}(R)]^2 \cos\theta_1 \cos\phi_1 \frac {\sin\phi_2} {\sin\theta_2} s_{nl}^{\theta \phi} =
f^{\theta \phi}_{62}(\vec a_{12}) \frac {d^2P_l(\cos\Delta)} {d\Delta^2} + f^{\theta \phi}_{61}(\vec a_{12}) \frac {dP_l(\cos\Delta)} {d\Delta}
\label{eq72}
\end{equation}

\begin{equation}
s^7_{nl} = - \beta_{nl}(R) \frac {\sin\phi_1} {\sin\theta_1}\sin\theta_2 \cos\phi_2 s_{nl}^{\phi r} =
f^{\phi r}_{71}(\vec a_{12}) \frac {dP_l(\cos\Delta)} {d\Delta}
\label{eq73}
\end{equation}

\begin{equation}
s^8_{nl} = - [\beta_{nl}(R)]^2 \frac {\sin\phi_1} {\sin\theta_1} \cos\theta_2 \cos\phi_2 s_{nl}^{\phi \theta} =
f^{\phi \theta}_{82}(\vec a_{12}) \frac {d^2P_l(\cos\Delta)} {d\Delta^2} + f^{\phi \theta}_{81}(\vec a_{12}) \frac {dP_l(\cos\Delta)} {d\Delta}
\label{eq74}
\end{equation}

\begin{equation}
s^9_{nl} = [\beta_{nl}(R)]^2 \frac {\sin\phi_1} {\sin\theta_1} \frac {\sin\phi_2} {\sin\theta_2} s_{nl}^{\phi \phi} =
f^{\phi \phi}_{92}(\vec a_{12}) \frac {d^2P_l(\cos\Delta)} {d\Delta^2} + f^{\phi \phi}_{91}(\vec a_{12}) \frac {dP_l(\cos\Delta)} {d\Delta}
\label{eq75}
\end{equation}

Where we evaluate the derivatives of $\Delta$
with respect to $\theta_1$, $\theta_2$, $\phi_1$ and $\phi_2$ using equations (\ref{eq54})-(\ref{eq57}) and the equations from \citep{winch-roberts95},
we obtain the following expressions

\begin{equation}
f^{rr}_{10}(\vec a_{12}) = N_l \sin \theta_1 \cos \phi_1 \sin \theta_2 \cos \phi_2
\label{eq76}
\end{equation}
\begin{equation}
f^{r \theta}_{21}(\vec a_{12}) = N_l \beta_{nl}(R) \sin \theta_1 \cos \phi_1 \cos \theta_2 \cos \phi_2 \cos \gamma_2
\label{eq77}
\end{equation}
\begin{equation}
f^{r \phi}_{31}(\vec a_{12}) = N_l \beta_{nl}(R) \sin \theta_1 \cos \phi_1 \sin \phi_2 \sin \gamma_2
\label{eq78}
\end{equation}
\begin{equation}
f^{\theta r}_{41}(\vec a_{12}) = N_l \beta_{nl}(R) \cos \theta_1 \cos \phi_1 \sin \theta_2 \cos \phi_2 \cos \gamma_1
\label{eq79}
\end{equation}
\begin{equation}
f^{\theta \theta}_{51}(\vec a_{12}) = - N_l [\beta_{nl}(R)]^2 \cos \theta_1 \cos \phi_1 \cos \theta_2 \cos \phi_2 \frac{\sin \gamma_1 \sin \gamma_2}
{\sin \Delta}
\label{eq80}
\end{equation}
\begin{equation}
f^{\theta \theta}_{52}(\vec a_{12}) = N_l [\beta_{nl}(R)]^2 \cos \theta_1 \cos \phi_1 \cos \theta_2 \cos \phi_2 \cos \gamma_1 \cos \gamma_2
\label{eq81}
\end{equation}
\begin{equation}
f^{\theta \phi}_{61}(\vec a_{12}) = N_l [\beta_{nl}(R)]^2 \cos \theta_1 \cos \phi_1 \sin \phi_2  \frac{\sin \gamma_1 \cos \gamma_2}{\sin \Delta}
\label{eq82}
\end{equation}
\begin{equation}
f^{\theta \phi}_{62}(\vec a_{12}) = N_l [\beta_{nl}(R)]^2  \cos \theta_1 \cos \phi_1 \sin \phi_2  \cos \gamma_1 \sin \gamma_2
\label{eq83}
\end{equation}
\begin{equation}
f^{\phi r}_{71} = - N_l \beta_{nl}(R) \sin \phi_1 \sin \theta_2 \cos \phi_2 \sin \gamma_1
\label{eq84}
\end{equation}
\begin{equation}
f^{\phi \theta}_{81}(\vec a_{12}) = - N_l [\beta_{nl}(R)]^2 \sin \phi_1 \cos \theta_2 \cos \phi_2 \frac{\cos \gamma_1 \sin \gamma_2}{\sin \Delta}
\label{eq85}
\end{equation}
\begin{equation}
f^{\phi \theta}_{82}(\vec a_{12}) = - N_l [\beta_{nl}(R)]^2 \sin \phi_1 \cos \theta_2 \cos \phi_2 \sin \gamma_1 \cos \gamma_2
\label{eq86}
\end{equation}
\begin{equation}
f^{\phi \phi}_{91}(\vec a_{12}) = N_l [\beta_{nl}(R)]^2 \sin \phi_1 \sin \phi_2 \frac{\cos \gamma_1 \cos \gamma_2}{\sin \Delta}
\label{eq87}
\end{equation}
\begin{equation}
f^{\phi \phi}_{92}(\vec a_{12}) = - N_l [\beta_{nl}(R)]^2 \sin \phi_1 \sin \phi_2  \sin \gamma_1 \sin \gamma_2
\label{eq88}
\end{equation}

In computing the cross correlation in equation (\ref{eq37}), we need to evaluate $d_{nl}^x(\vec R_1, \vec R_2)$
in equation (\ref{eq34}) as
\begin{equation}
d_{nl}^x(\vec R_1, \vec R_2) =  \sum_{m} d_{nlm}^x(\vec R_1)d_{nlm}^x(\vec R_2) =
\sum_{i=1}^9 s^i_{nl} = f_0 P_l(\cos\Delta) + f_1 \frac {dP_l(\cos\Delta)} {d\Delta}
+ f_2 \frac {d^2P_l(\cos\Delta)} {d\Delta^2}
\label{eq89}
\end{equation}
where,
\begin{equation}
f_0(\vec a_{12}) = f^{rr}_{10}
\label{eq90}
\end{equation}
\begin{equation}
f_1(\vec a_{12}) = f^{r \theta}_{21} + f^{r \phi}_{31} + f^{\theta r}_{41} + f^{\theta \theta}_{51} + f^{\theta \phi}_{61} +
f^{\phi r}_{71} + f^{\phi \theta}_{81} + f^{\phi \phi}_{91}
\label{eq91}
\end{equation}
\begin{equation}
f_2(\vec a_{12}) = f^{\theta \theta}_{52} + f^{\theta \phi}_{62} + f^{\phi \theta}_{82} + f^{\phi \phi}_{92}
\label{eq92}
\end{equation}

\section{Appendix D}

In this appendix we derive the angular coordinates $(\theta_2, \phi_2)$
of point B on the annulus, which is at a fixed annular angle $\gamma_1$
and a given angular distance $\Delta$ from the fixed point A with
 angular coordinates $(\theta_1, \phi_1)$ at the center of the annulus.
We use equations (\ref{eq53}) and (\ref{eq54}) from Appendix C
to express the angular coordinates of B in terms of the fixed angles $(\theta_1, \phi_1, \gamma_1, \Delta)$.
The geometry for the calculation is shown in Figure 6.

\begin{equation}
\cos \Delta = \cos \theta_1 \cos \theta_2 + \sin \theta_1 \sin \theta_2 \cos(\phi_1 - \phi_2)
\label{eq93}
\end{equation}
\begin{equation}
\sin \Delta \cos \gamma_1 = \sin \theta_1 \cos \theta_2 - \cos \theta_1 \sin \theta_2 \cos(\phi_1 - \phi_2)
\label{eq94}
\end{equation}

We now consider the case when the two points are close together, which is true for small distances $\Delta$.
They are related as $\theta_2 = \theta_1 + \delta \theta_1$ and $\phi_2 = \phi_1 + \delta \phi_1$, for
small increments $\delta \theta_1$ and $\delta \phi_1$. The  equations (\ref{eq93}) and (\ref{eq94}) can be expanded in terms of the increments,
and using the standard trigonometric approximations for small angles $\sin \theta \approx \theta$ and
$\cos \theta \approx 1 - \frac{\theta^2}{2}$ we obtain,

\begin{equation}
\delta \theta_1 = \Delta \left[1 - \frac{\sin^2 \gamma_1 \sin^2 \theta_1}{\sin^2 \theta_1 - \frac{\Delta}{2} \cos \gamma_1 \sin 2\theta_1} \right]^{\frac{1}{2}}
\label{eq95}
\end{equation}
\begin{equation}
\delta \phi_1 = \frac{\Delta \sin \gamma_1}{ \left[\sin^2 \theta_1 - \frac{\Delta}{2} \cos \gamma_1 \sin 2\theta_1 \right]^{\frac{1}{2}}}
\label{eq96}
\end{equation}

For fixed $\Delta$ and $\gamma_1$
we use equations (\ref{eq95}) and (\ref{eq96}) to calculate  $(\theta_2, \phi_2)$ for small angular separation.
For larger separations we can solve the linear equations (\ref{eq93}) and (\ref{eq94}) to calculate $\theta_2$, and equation
(\ref{eq56}) to compute $\phi_2$, and equations (\ref{eq55}) and (\ref{eq57}) to calculate $\cos \gamma_2$ and $\sin \gamma_2$ respectively.

\section{Appendix E}

\begin{figure}[!ht]
\epsscale{0.8} \plotone{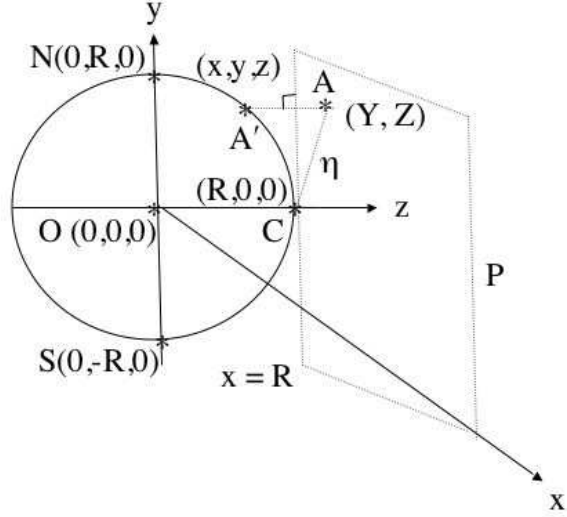} \caption{Orthographic projection.
The point C is the disk center and $\eta$ measures the distance from
C to any projected point (Y,Z) on the orthographic plane P, which is
at $x = R$. Any point A$^{'}$ on the solar surface is specified by
$(\theta_1, \phi_1)$ or
the angle $\phi_1$ and the distance $\eta$. } \label{fg7}
\end{figure}

\begin{figure}[!ht]
\epsscale{0.8} \plotone{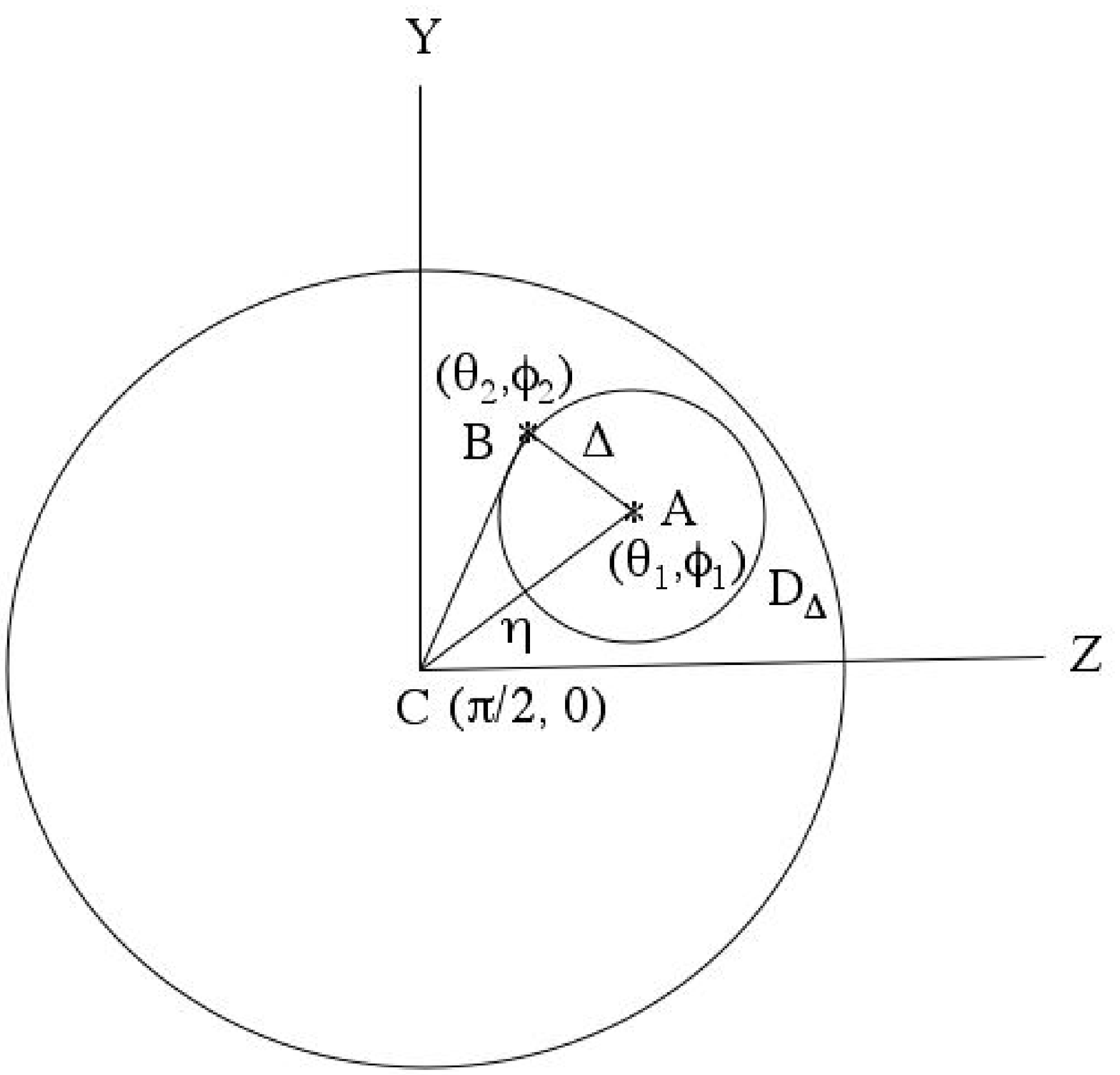} \caption{Projected Disc. This shows
the annulus $D_{\Delta}$ of radius $\Delta$ around the point A
$(\theta_1, \phi_1)$ on the solar surface. The cross correlation is
computed between the annulus center A  and a point B $(\theta_2,
\phi_2)$ on the solar surface, that has an angular  separation of
$\Delta$ from A. The point A is at a distance $\eta$ from the disk
center C, $(\theta_c, \phi_c) = (\frac{\pi}{2}, 0)$. } \label{fg8}
\end{figure}

In this appendix we consider the orthographic projection shown in
Figure~\ref{fg7}, where the points on the surface of the Sun are
projected onto the $YZ$ plane P, which is perpendicular to the line of
sight direction $\hat e_x$. The origin O is at the center of the Sun
of radius $R$. We denote a point A$^{'}$ on the solar surface as
$(x, y, z)$ with axis $xyz$ and as $(Y, Z)$ the coordinate of the same
point projected as A on the projection plane P with axis $YZ$. The $x$
axis is perpendicular the the projection plane. The projection is
oriented with north along  $+y$ axis and east along the $+z$ axis.
The projection is centered at co-latitude of $90^\circ$ and
longitude of $0^\circ$. With this convention, the north pole N has
coordinates $(0, R, 0)$, the south pole S is $(0, -R, 0)$, and the
point of tangency C between the sphere and the plane is $(R, 0, 0)$.
The point C is also called the disk center. The projection plane is
$x = R$. The coordinates of any point A$^{'}$ with co-latitude
$\theta_1$ and longitude $\phi_1$ on the solar surface are
\begin{equation}
(x, y, z) = (R \sin \theta_1 \cos \phi_1, R \sin \theta_1 \sin \phi_1, R \cos \theta_1 )
\label{eq97}
\end{equation}
The equation of the sphere is $x^2 + y^2 + z^2 = R^2$. The
coordinates of the same point on the plane are  $(Y, Z) = (y, z)$.
The point of tangency C  between the sphere and the plane is the
disk center and has coordinates $(Y, Z) = (0, 0)$ on the plane.
Hence, the distance from the disk center C $(0, 0)$ to any point A
$(Y, Z)$ on the plane P is denoted by $\eta$, which is fixed for
each computation of the cross correlation. Any point A$^{'}$ on the
solar surface and hence the corresponding projection A on the plane
P can be specified by $(\theta_1, \phi_1)$ or alternatively by the two parameters  $\eta$ and the angle
$\phi_1$. In  Figure~\ref{fg8}, the center to annulus geometry is
shown with the fixed center point A $(\theta_1, \phi_1)$, which is
at a distance $\eta$ from disk center C $(Y, Z) = (0, 0)$, and a
point B $(\theta_2, \phi_2)$ is on the annulus $D_{\Delta}$ of
radius $\Delta$.
\begin{equation}
\eta = (Y^2 + Z^2)^{\frac{1}{2}} = (y^2 + z^2)^{\frac{1}{2}} = (R^2 - x^2)^{\frac{1}{2}} = R(1 - \sin^2 \theta_1 \cos^2 \phi_1)^{\frac{1}{2}}
\label{eq98}
\end{equation}

The angular coordinates of the disk center C, are $(\theta_c, \phi_c) = (\frac{\pi}{2}, 0)$. When the point A is at the disk center
$(\theta_1, \phi_1) = (90^\circ, 0^\circ)$, and hence the distance $\eta = 0$.
For fixed values of $(\Delta, \gamma_1)$, together with the coordinates $(\theta_1, \phi_1)$ of the annulus center A, we can compute the
coordinates $(\theta_2, \phi_2)$ of point B on the annulus by solving the linear system of equations  (\ref{eq93}) and (\ref{eq94})
to get $\theta_2$ and solving equation (\ref{eq56}) to compute  $\phi_2$.


\begin{thebibliography}{}
\bibitem[Bracewell(1999)]{bracewell-book} Bracewell, R., 2000, The Fourier Transform and Its
Applications, 3rd edition (New York: McGraw-Hill)
\bibitem[Christensen-Dalsgaard(2002)]{jcd02} Christensen-Dalsgaard, J. 2002, {\it Rev. Mod. Phys.}, 74, 1073
\bibitem[Christensen-Dalsgaard(2003)]{dalsgaardbook} Christensen-Dalsgaard, J. 2003, Lecture Notes on Stellar
Oscillations, 5th edition.
\bibitem[Duvall et al.(1993)]{duvalletal93} Duvall, T. L., Jr., Jefferies, S. M.,
Harvey, J. W. \& Pomerantz, M. A. 1993, \nat, 362, 430
\bibitem[Duvall et al.(1997)]{duvalletal97} Duvall, T. L., Jr. et al. 1997, Sol. Phys., 170, 63
\bibitem[Duvall(2003)]{duvall02} Duvall, T.~L., Jr.\ 2003, ESA
SP-517: GONG+ 2002.~Local and Global Helioseismology: the Present and
Future, 12, 259
\bibitem[Giles (1999)]{giles99} Giles, P. M. 1999, Ph.D. Thesis, Stanford University
\bibitem[Gradshteyn and Ryzhik(1994)]{grad-ryzhik94} Gradshteyn, I. S. and Ryzhik, I. M. 1994, page 531,
Table of Integrals, Series, and Products (5th edition), Academic Press, San Diego
\bibitem[Jackson(1999)]{jackson-book} Jackson, J. D., 1999, Classical Electrodynamics, 3rd edition
(New York: John Wiley \& Sons)
\bibitem[Jefferies et al.(1994)]{jefferiesetal94} Jefferies, S.M., Osaki, Y., Shibahashi, H., Duvall, T.L., Jr.,
Harvey, J.W., \& Pomerantz, M.A. 1994, \apj, 434, 795
\bibitem[Kosovichev(1996)]{kosovichev1996} Kosovichev, A.~G.\ 1996, \apjl, 461, L55
\bibitem[Kosovichev \& Duvall(1997)]{sashaduvall97} Kosovichev, A.G., Duvall, T.L., Jr. in SCORe-'96:
Solar Convection and Oscillations and Their Relationship, ed. F.P. Pijpers et al., 1997 (Kluwer)
\bibitem[Lighthill(1978)]{lighthill-book} Lighthill, M.J., 1978, Waves in Fluids (Cambridge: Cambridge University Press)
\bibitem[Smart(1977)]{smart-book} Smart, W. M., 1977, Textbook on Spherical Astronomy, 6th edition (Cambridge: Cambridge University Press)
\bibitem[Unno et al.(1989)]{unno-book} Unno, W., Osaki, Y., Ando, H., \& Shibahashi, H., 1989, Nonradial Oscillations of Stars
(University of Tokyo Press, Tokyo)
\bibitem[Whitham(1974)]{whitham-book} Whitham, G.B., 1974, Linear and Nonlinear Waves
(New York: John Wiley \& Sons)
\bibitem[Winch and Roberts(1995)]{winch-roberts95} Winch, D. E. and Roberts, P. H. 1995, J. Austral Math Soc Ser B 37, 212
\bibitem[Zhao(2004)]{zhao-thes} Zhao, J. 2004 Ph.D. Thesis, Stanford University
\end{thebibliography}
\end{document}